\documentclass{article}

\begin{document}
\begin{center}
{\Large\bf 
Algebraic structure of the Feynman propagator and a new 
correspondence for canonical transformations 
}
\vspace{5cm}

Akihiro Ogura and Motoo Sekiguchi${}^{\dagger}$
 
\vspace{1cm}

Laboratory of Physics, Nihon University, Matsudo 271-8587, Japan

and 

${}^{\dagger}$Faculty of Engineering, Kokushikan University, 
Tokyo 154-8515, Japan

\vspace{1.5cm}

\begin{abstract}
We investigate the algebraic structure of the Feynman propagator 
with a general time-dependent quadratic Hamiltonian system. Using 
the Lie-algebraic technique we obtain a normal-ordered form of 
the time-evolution operator, and then the propagator is easily 
derived by a simple ``Integration Within Ordered Product" (IWOP) 
technique.It is found that this propagator contains a classical 
generating function which demonstrates a new correspondence 
between classical and quantum mechanics.

\begin{flushleft}
\end{flushleft}
\end{abstract}
\end{center}

\newpage
\section{Introduction}

In recent years there has been considerable interest in the 
application of the Lie-algebraic technique in order to derive 
the propagators or density matrices of the systems. These systems 
include not only a free particle system but also a harmonic 
oscillator~\cite{wang}, a charged oscillator in a constant magnetic 
field~\cite{yonei1}, and a general time-dependent oscillator~\cite{lo}. 
In those papers cited, the direct use of the position and momentum 
operators plays an important role. 
However, creation and annihilation operators are often used for 
representation of the Lie algebra and applied to harmonic 
oscillator systems. It is recognized~\cite{yonei2} that the use of 
creation and annihilation operators clarifies the physics and 
simplifies the calculation. 

The purpose of this paper is to present a method in which the Lie 
algebra of the squeezed operator plays a direct role in deriving 
the propagator of a general time-dependent quadratic Hamiltonian 
system. It is found that this propagator contains a classical 
generating function which implies a new correspondence between 
classical and quantum mechanics.
In carrying out this program, we focus on a time-evolution operator 
with a general time-dependent quadratic Hamiltonian system as follows: 
\begin{equation}
U(t)=\exp\left[ - \frac{i}{2} 
     \left\{ \alpha(t) p^{2} 
              + \beta(t) \left( qp+pq \right) 
            + \gamma(t) q^{2} \right\}\right], 
\label{eq:uni1}
\end{equation}
where $q$ and $p$ are the coordinate and momentum variables and 
$\alpha(t)$, $\beta(t)$, and $\gamma(t)$ are arbitrary real 
functions of time $t$. 
In the next section we first review squeezed operators and then 
calculate the propagator $<Q|U(t)|q>$ with the aid of 
the ``Integration Within Ordered Product" (IWOP) technique. 
In this case the creation and annihilation operators play a central 
role in deriving the propagator. 
In section 3, we investigate the character of the propagator from 
the point of linear transformations in the coordinate-momentum 
phase space. We then show a new correspondence between classical 
and quantum mechanics. 
Section 4 is devoted to a discussion.

\section{Squeezed operator and the propagator}

The squeezed operators~\cite{stoler, yuen} are defined using 
the annihilation operator $a=(q+ip)/\sqrt{2}$ as follows: 
\begin{equation}
K_{+} = \frac{{a}^{\dagger} a^{\dagger}}{2}, \hspace{5mm}
K_{0} = \frac{a^{\dagger} a}{2}+\frac{1}{4}, \hspace{5mm}
K_{-} = \frac{a a}{2}, 
\label{eq:squeez}
\end{equation}
which form the SU(1, 1) Lie algebra 
\begin{equation}
\left[ K_{+}, K_{-} \right] = - 2K_{0}, \hspace{5mm}
\left[ K_{0}, K_{\pm} \right] = \pm K_{\pm} .
\end{equation}

We rewrite the unitary operator (\ref{eq:uni1}) in terms of the 
squeezed operators (\ref{eq:squeez}) 
\begin{equation}
U(t) = \exp\left[ \tau(t) K_{+} + i \sigma(t) K_{0} 
                 -\tau^{\ast}(t) K_{-} \right], 
\label{eq:uni2}
\end{equation}
where 
\begin{equation}
\left\{
\begin{array}{ll}
\tau(t)   &= \beta(t)+i\frac{\alpha(t)-\gamma(t)}{2} \\
\sigma(t) &= -\alpha(t)-\gamma(t) ,
\end{array}
\right.
\label{eq:coef1}
\end{equation}
$\tau(t)$ is a complex function and $\sigma(t)$ is a 
real function of time $t$. 
Using the technique of differential equations~\cite{truax, fisher}, 
we obtain the normal ordered form of the operator (\ref{eq:uni2}) 
\begin{equation}
U(t) = \exp\left[- \frac{r(t)}{s(t)} K_{+} \right]
       \exp\left[- 2 K_{0} \ln s(t) \right]
       \exp\left[\frac{r^{\ast}(t)}{s(t)} K_{-} \right], 
\label{eq:uni3}
\end{equation}
where $s(t)$ and $r(t)$ are defined by 
\begin{equation}
\left\{
\begin{array}{ll}
s(t) &= \cosh \Delta - \frac{i\sigma(t)}{2\Delta}\sinh\Delta \\
r(t) &= -\frac{\tau(t)}{\Delta}\sinh\Delta
\end{array}
\right.
\label{eq:coef2}
\end{equation}
with 
\begin{equation}
\Delta^{2} = |\tau(t)|^{2}-\frac{\sigma^{2}(t)}{4} 
           = \beta^{2}(t)-\alpha(t) \gamma(t)
\end{equation}
which satisfy the following relation; 
\begin{equation}
|s(t)|^{2}-|r(t)|^{2} = \cosh^{2} \Delta - \sinh^{2} \Delta = 1 .
\end{equation}

Now, we are in a position to calculate the Feynman propagator 
$<Q|U(t)|q>$. 
The coherent state $|z>$ is defined~\cite{zhang} by the eigenstate of 
the annihilation operator $a$ with the complex eigenvalue $z$, i.e. 
\begin{equation}
a|z> = z|z>, 
\end{equation}
and form the completeness relation 
\begin{equation}
\int \frac{d^{2}z}{2\pi i} |z><z| = 1, 
\label{eq:comp}
\end{equation}
where $\int d^{2}z \equiv \int d[{\rm Re}(z)] d[{\rm Im}(z)]$. 
To obtain the Feynman propagator $<Q|U(t)|q>$, 
we use the completeness relation (\ref{eq:comp}) with 
arguments of $z_{1}$ and $z_{2}$ 
\begin{equation}
<Q|U(t)|q>=\int\frac{d^{2}z_{1} d^{2}z_{2}}{(2\pi i)^{2}}
<Q|z_{1}><z_{1}|U(t)|z_{2}><z_{2}|q>.
\label{eq:kernel1}
\end{equation}
With the aid of the IWOP technique~\cite{fan1, fan2}, 
\begin{equation}
<z_{1}|U(t)|z_{2}> = \frac{1}{\sqrt{s}}
\exp\left[ -\frac{r}{2s}(z_{1}^{\ast})^{2}
+\frac{z_{2} z_{1}^{\ast}}{s}+\frac{r^{\ast}}{2s}z_{2}^{2}
-\frac{|z_{1}|^{2}}{2}-\frac{|z_{2}|^{2}}{2}
 \right]
\end{equation}
and the coherent state with coordinate representation 
\begin{eqnarray}
<z_{2}|q> &=& \frac{1}{\pi^{1/4}} 
\exp\left[ -\frac{q^{2}}{2}+\sqrt{2}qz_{2}^{\ast}
-\frac{(z_{2}^{\ast})^{2}}{2}-\frac{|z_{2}|^{2}}{2} \right], \\
<z_{1}|Q> &=& \frac{1}{\pi^{1/4}} 
\exp\left[ -\frac{Q^{2}}{2}+\sqrt{2}Qz_{1}^{\ast}
-\frac{(z_{1}^{\ast})^{2}}{2}-\frac{|z_{1}|^{2}}{2} \right], 
\end{eqnarray}
we integrate $z_{1}$ and $z_{2}$ of (\ref{eq:kernel1}) to obtain 
\begin{eqnarray}
<Q|U(t)|q> &=& \sqrt{\frac{1}{\pi(s-s^{\ast}-r+r^{\ast})}}
                \times \nonumber \\
& & \exp\left[ \frac{2qQ}{s-s^{\ast}-r+r^{\ast}}
  -\frac{q^{2}}{2}\frac{s+s^{\ast}-r-r^{\ast}}{s-s^{\ast}-r+r^{\ast}}
  -\frac{Q^{2}}{2}\frac{s+s^{\ast}+r+r^{\ast}}{s-s^{\ast}-r+r^{\ast}}
 \right]. 
\label{eq:kernel2}
\end{eqnarray}
This is the transition amplitude $q \to Q$ derived from 
the unitary operator (\ref{eq:uni1}).

\section{Propagator and linear canonical transformations}

In this section, we shall investigate the character of 
the propagator (\ref{eq:kernel2}). 
Before doing so, we shall concentrate on the meaning of the operator 
(\ref{eq:uni1}) in terms of the coordinate-momentum phase space. 
Transforming the annihilation and creation operators under 
(\ref{eq:uni3}), we obtain the following form for the operator: 
\begin{equation}
\left\{
\begin{array}{ll}
U^{\dagger}(t)aU(t) &=  s^{\ast}(t) a-r(t) a^{\dagger} \\
U^{\dagger}(t)a^{\dagger}U(t) &=  s(t) a^{\dagger}-r^{\ast}(t) a, 
\end{array}
\right.
\label{eq:linear1}
\end{equation}
which is described in the coordinate-momentum phase space 
\begin{equation}
\left\{
\begin{array}{ll}
Q(t) &= U^{\dagger}(t) q U(t) =  A(t) q + B(t) p \\
P(t) &= U^{\dagger}(t) p U(t) =  C(t) q + D(t) p, 
\end{array}
\right.
\label{eq:linear2}
\end{equation}
where $(Q, P)$ and $(q, p)$ are the new and old quantum canonical 
variables which combine linearly with the time-dependent 
real functions $A(t)$, $B(t)$, $C(t)$, and $D(t)$. 
Also, these coefficients satisfy 
\begin{equation}
\left\{
\begin{array}{ll}
s(t) &= \frac{D(t)+A(t)}{2}+i\frac{B(t)-C(t)}{2} \\
r(t) &= \frac{D(t)-A(t)}{2}-i\frac{B(t)+C(t)}{2}, 
\end{array}
\right.
\label{eq:coef3}
\end{equation}
subject to 
\begin{equation}
|s(t)|^{2}-|r(t)|^{2} = A(t)D(t)-B(t)C(t) = 1, 
\label{eq:abcd}
\end{equation}
which signifies the existense of the inverse of the linear 
transformation (\ref{eq:linear2}). 
On the other hand, by straightforward calculation 
$U^{\dagger}(t) q U(t)$ and $U^{\dagger}(t) p U(t)$ with the operator
(\ref{eq:uni1}) using the Baker-Campbell-Hausdorff formulae, 
we can assign 
the coefficients $A(t)$, $B(t)$, $C(t)$, and $D(t)$ in terms of 
$\alpha(t)$, $\beta(t)$, and $\gamma(t)$ as follows: 
\begin{eqnarray}
\left(\begin{array}{cc}
        A(t) & B(t) \\
        C(t) & D(t)
      \end{array}
\right) 
&=&
\left(\begin{array}{cc}
        \cosh \Delta + \frac{\beta(t)}{\Delta} \sinh \Delta 
        & \frac{\alpha(t)}{\Delta} \sinh \Delta \\
        - \frac{\gamma(t)}{\Delta} \sinh \Delta
        & \cosh \Delta-\frac{\beta(t)}{\Delta} \sinh \Delta
      \end{array}
\right) \label{eq:coef4} \\
&=& 
\left(\begin{array}{cc}
          \frac{s+s^{\ast}-r-r^{\ast}}{2}
        & \frac{-is+is^{\ast}+ir-ir^{\ast}}{2} \\
          \frac{is-is^{\ast}+ir-ir^{\ast}}{2}
        & \frac{s+s^{\ast}+r+r^{\ast}}{2}
      \end{array}
\right).  
\label{eq:coef5}
\end{eqnarray}
Note that these coefficients (\ref{eq:coef4}) and (\ref{eq:coef5}) 
are  consistent with (\ref{eq:coef2}) and (\ref{eq:coef3}).

Now we rewrite the propagator (\ref{eq:kernel2}) in terms of 
$A(t)$, $B(t)$, $C(t)$, and $D(t)$ from (\ref{eq:coef5}), then 
\begin{equation}
<Q|U(t)|q> = \sqrt{\frac{1}{2\pi i B(t)}}
             \exp\left[ -i W(q, Q, t) \right], 
\label{eq:kernel3}
\end{equation}
where 
\begin{equation}
W(q, Q, t) = \frac{qQ}{B(t)} - \frac{A(t)}{2B(t)}q^{2} 
                             - \frac{D(t)}{2B(t)}Q^{2}. 
\label{eq:gfunction}
\end{equation}
This propagator is for the most general time-dependent quadratic 
Hamiltonian system, 
and thus results for any spacial case can easily be deduced from it. 
For example, we take the harmonic oscillator. In this case, we have 
\begin{equation}
\alpha(t)=\frac{t}{m}, \hspace{5mm} \beta(t)=0, \hspace{5mm} 
\gamma(t)=m \omega^{2}t, 
\end{equation}
then $\Delta=i \omega t$ and so 
\begin{equation}
\left(\begin{array}{cc}
        A(t) & B(t) \\
        C(t) & D(t)
      \end{array}
\right)=
\left(\begin{array}{cc}
        \cos\omega t & \frac{1}{m\omega}\sin \omega t \\
        -m\omega \sin\omega t & \cos \omega t
      \end{array}
\right)
\end{equation}
We recover the well known result for the Feynman propagator 
\begin{equation}
<Q|U(t)|q>=\sqrt{\frac{m\omega}{2\pi i \sin\omega t}}
\exp\left[i\left\{
-\frac{m\omega}{\sin\omega t}qQ+\frac{m\omega}{2\tan\omega t}
\left( q^{2}+Q^{2} \right)
\right\}
\right]
\end{equation}
In the limit $\omega \to 0$, we fall back to the free particle 
case; i.e. $\alpha(t)=\frac{t}{m}, \beta(t)=\gamma(t)=0$. Then, 
we obtain 
\begin{equation}
\left(\begin{array}{cc}
        A(t) & B(t) \\
        C(t) & D(t)
      \end{array}
\right)=
\left(\begin{array}{cc}
        1 & \frac{t}{m} \\
        0 & 1
      \end{array}
\right)
\end{equation}
and
\begin{equation}
<Q|U(t)|q>=\sqrt{\frac{m}{2\pi i t}}
\exp\left[i\left\{
-\frac{m}{t}qQ+\frac{m}{2t}
\left( q^{2}+Q^{2} \right)
\right\}
\right]. 
\end{equation}

It is worth mentioning two points here. The first point is that
the exponential function $W(q, Q, t)$ in (\ref{eq:gfunction}) is 
a generating function that gives rise to a classical linear canonical 
transformation (\ref{eq:linear2}) with an ordinary prescription in 
classical mechanics; 
\begin{equation}
p=\frac{\partial W}{\partial q}, \hspace{5mm} 
P=- \frac{\partial W}{\partial Q} .
\end{equation}
It was Dirac who first discussed that the exponential of the classical 
generating function can be used as the quantum transformation 
function~\cite{dirac1, dirac2, dirac3}. This remark lead Feynman 
to his path integral formulation of quantum mechanics~\cite{feynman}. 
The second point is that we have derived the propagator 
(\ref{eq:kernel3}) with a general time-dependent quadratic Hamiltonian 
system (\ref{eq:uni1}), which reflects a linear transformation 
in coordinate-momentum phase space (\ref{eq:linear2}). 
Let us consider the reverse, once we have the linear transformation 
(\ref{eq:linear2}) in coordinate-momentum phase space, the propagator 
is written down at once in the form (\ref{eq:kernel3}). 
These two points reveal a new correspondence between classical 
and quantum mechanics. 

\section{Summary}

We have presented a method by which the propagator of a general 
time-dependent quadratic Hamiltonian system can be derived by the 
Lie-algebraic technique for the squeezed operators. 
It was found that this propagator contains a generating function 
which gives rise to a linear transformation in coordinate-momentum 
phase space in classical mechanics. 
Furthermore, our formulation has a useful attribute in that 
once we have the linear canonical transformation in 
coordinate-momentum phase space, its 
quantum counterpart of a unitary operator and the Feynman propagator 
are easily calculated.  

\vspace{1cm}


\end{document}